\documentclass[a4paper]{jpconf}

\usepackage{graphicx}
\newcommand{\diff}{\mathrm d}
\renewcommand{\v}[1]{{\bf #1}}

\begin{document}

\title{Molecular Dynamics Study of Rotating Nanodroplets: Finite-size Effects and Nonequilibrium Deformation}

\author{ Hiroshi Watanabe$^{1,2}$, Naoki Mitsuda$^3$, Tomoaki Nogawa$^3$, and  Nobuyasu Ito$^3$}

\address{
$^1$ Institute for Solid State Physics, The University of Tokyo,
Kashiwanoha 5-1-5, Kashiwa, Chiba 277-8581, JAPAN \\
$^2$ Super Computing Division
Information Technology Center
University of Tokyo, 2-11-16 Yayoi, Bunkyo, Tokyo 113-8658, Japan\\
$^3$ Department of Applied Physics, School of Engineering,
The University of Tokyo, Hongo, Bunkyo-ku, Tokyo 113-8656, Japan
}
\ead{hwatanabe@issp.u-tokyo.ac.jp}
\begin{abstract}
Noneqiuilibrium dynamics of rotating droplets are studied by molecular dynamics simulations.
Small deviations from the theoretical prediction are observed when the size of a droplet is small,
and the deviations become smaller as the size of the droplet increases.
The characteristic timescale of the deformation is observed, and we find
(i) the deformation timescale is almost independent of the rotating velocity
with for small frequency and
(ii) the deformation timescale becomes shorter as temperature increases.
A simple model is proposed to explain the deformation dynamics of droplets.
\end{abstract}

\section{Introduction}

Droplets are small drops of liquid which are commonly observed in our life.
Many things can be considered as ensembles of liquid droplets, 
such as rain, clouds, and splays, etc.
Droplets are not only familiar to us, but their behaviors are also important for engineering,
such as spray combustion~\cite{spray_combustion},
developments of inkjet printers~\cite{inkjet} and electronic sputtering~\cite{sputtering}, etc.
Behaviors of droplets are mainly governed by two kinds of forces,
namely, the surface tension and the inertial force. The surface tension works as
restoring force trying to keep shape of a droplet, while the inertial force
usually tries to deform and destroy it.
A simple example of such balance between two forces
can be seen in rotating droplets where the centrifugal force plays the role of the inertial force.
One of the fundamental studies of such rotating droplets was investigated by Plateau~\cite{Plateau}.
A water-droplet was put in alcohol with same density in order to mimic a gravity-less system,
and then the container was rotated.
The droplet was gradually compressed to flat shape as the rotating velocity increased
and became unstable when the angular velocity exceeded some critical value.
While the experimental technique is quite simple,
it is difficult to investigate this phenomena analytically because of the influence from the ambient fluid.
Recently, levitated droplets have attracted much researchers' interests.
A droplet is levitated by some external force, such as
electromagnetic force~\cite{PhysRevLett.101.234501}, and then a containerless system is achieved
which is convenient for comparison with theoretical predictions.
The  properties of rotating droplets in steady-state are investigated theoretically,
and numerical works were followed~\cite{Brown, Watanabe2009867}.
Also, the gravityless experiments were performed in spacelab by Wang \textit{et al.}~\cite{PhysRevLett.56.452}. They reported that while the axisymmetric shapes are well described by the 
theoretical prediction, the bifurcation point from axisymmetric to nonaxisymmetric shapes
locates at a lower rotation velocity than the theoretical prediction.
Despite of such past studies, dynamics of the deformation, especially nonlinear motions
which cannot be described as simple oscillations, have not been clarified yet.
Dynamical aspects of the deformation, such as characteristic timescale of the deformation or
how the droplets separates into fragmentations, are very 
important both for theoretical interests and for applications.
Additionally, the validity of the continuum treatments should be tested for small droplets,
since a droplet consists of many particles, and consequently,
the surface of it has finite thickness, while thickness is usually ignored in the continuum theory.
In the present study, we investigate the dynamics of the rotating droplets
using molecular dynamics (MD) simulations. Using MD, dynamics of droplets
can be studied naturally taking account of its thickness.
We first make a brief review of the continuum treatments for rotating droplets in the steady state,
and compare it to our results after describing details of numerical methods.
Deformation dynamics are observed for different conditions, and a simple model is
proposed in order to explain the results.

\begin{figure}[tb]
\begin{center}
\includegraphics[width=0.5\linewidth]{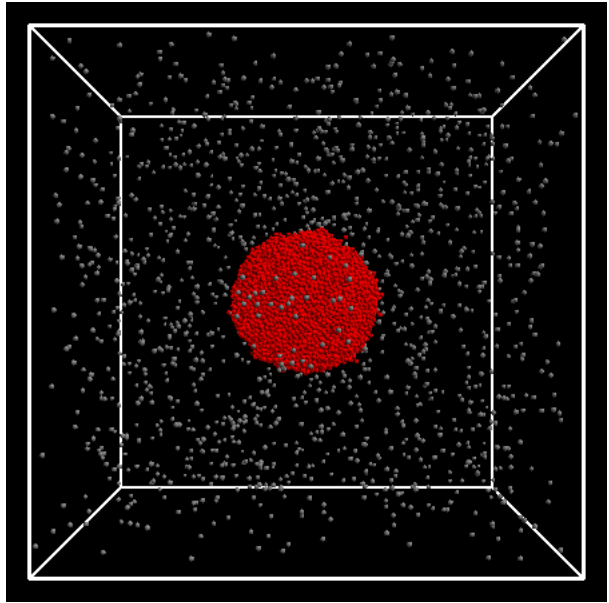}
\caption{A typical snapshot of a static droplet.
The white lines denotes the simulation box with the linear size $L=40$.
The total number of particles $N=16000$ and the temperature $T=0.6$.
The red and gray particles denote a liquid droplet and vapor, respectively.
}
\label{fig::droplet}
\end{center}
\end{figure}

\section{Form of Rotating Droplet} \label{sec::theory}

The form of rotating droplets are solved exactly by Chandrasekhar~\cite{Chandrasekhar} when its shape is uniaxial.
The steady forms of rotating droplets are determined by the balance
between the surface tension and the centrifugal force.
Here, we make the brief review of the form of the rotating droplet 
following the manner by Chandrasekhar.

Consider a droplet rotating around $z$-axis in a system without the gravity.
In the steady state, the gradient of the internal pressure of the droplet should be 
balanced by the centrifugal force as
\begin{equation}
 \rho \omega^2 r = \frac{\diff p}{\diff r}, \label{eq::pressure}
\end{equation}
where $\rho$ is the density of the droplet, $\omega$ is angular velocity,
$p$ is the internal pressure, and $r$ is the distance from the rotating axis, respectively.
Equation~(\ref{eq::pressure}) directly leads to 
\begin{equation}
p(r) = p_0 + \frac{1}{2} \rho \omega^2 r^2, \label{eq::pressure2}
\end{equation}
with the pressure $p_0$ at the center of the droplet ($r=0$).
At the surface of the droplet, the pressure should satisfy the following relation
\begin{equation}
p = \gamma \mathrm{div} \v{n}, \label{eq::gamma}
\end{equation}
with the surface tension $\gamma$ and the normal vector of the droplet surface $\v{n}$.
From Eqs.~(\ref{eq::pressure2}) and (\ref{eq::gamma}), we have
\begin{equation}
p_0 + \frac{1}{2} \rho \omega^2 r^2 = \gamma \mathrm{div} \v{n}. \label{eq::pressure_gamma}
\end{equation}
While shape of a rotating droplet varies from the sphere,
the shape will be uniaxial when the rotating speed is small.
Therefore, the form of the droplet can be expressed as $z = f(r)$,
where $z$ is the height from the equatorial plane and $r$ is distance from $z$-axis,
namely, $r^2 = x^2 + y^2$, respectively.
Then the unit normal vector of the droplet surface $\v{n} = (n_x,n_y,n_z)$ is
\begin{eqnarray}
n_x &=& - \frac{\phi}{\sqrt{1+\phi^2}} \frac{x}{r}, \\
n_y &=& - \frac{\phi}{\sqrt{1+\phi^2}} \frac{y}{r}, \\
n_z &=&  \frac{1}{\sqrt{1+\phi^2}},
\end{eqnarray}
where
\begin{equation}
\phi \equiv \frac{\diff f}{\diff r}.
\end{equation}
From the above, we easily have,
\begin{eqnarray}
\mathrm{div}{\v{n}} &=& \displaystyle \frac{\partial n_x}{\partial x} + \frac{\partial n_y}{\partial y} + \frac{\partial n_z}{\partial z}, \\
&=&  - \frac{\phi}{r(1+\phi^2)^{1/2}} - \frac{1}{(1+\phi^2)^{3/2}} \frac{\diff \phi}{\diff r}, \\
&=& - \frac{1}{r} \frac{\diff}{\diff r} \frac{r \phi}{\sqrt{1+\phi^2}}. \label{eq::divn}
\end{eqnarray}
Substituting Eq.~(\ref{eq::divn}) to Eq.~(\ref{eq::pressure_gamma}), we obtain
\begin{equation}
p_0 +\frac{1}{2} \rho \omega^2 r^2  =  - \frac{\gamma}{r} \frac{\diff}{\diff r} \frac{r \phi}{\sqrt{1+\phi^2}}, 
\end{equation}
which is integrable to be
\begin{equation}
\frac{p_0r^2}{2 \gamma}  +\frac{ \rho \omega^2 r^4}{8\gamma}  = - \frac{r \phi}{\sqrt{1+\phi^2}}. \label{eq::r_phi}
\end{equation}
Note that, the constant of the integration should be zero because of the rotational symmetry around $z$-axis.
Let $R_\mathrm{e}$ be the equatorial radius of the rotating droplet.
Since $\phi \rightarrow - \infty$ when $r\rightarrow R_\mathrm{e}$, we have
\begin{equation}
\frac{p_0 R_\mathrm{e}}{2 \gamma} + \frac{\rho \omega^2 R_\mathrm{e}^3}{8 \gamma} = 1.
\end{equation}
We introduce a dimensionless value $\Sigma$ as
\begin{equation}
\Sigma \equiv  \frac{\rho \omega^2 R_\mathrm{e}^3}{8 \gamma} \equiv \left(\frac{\omega}{\omega_0}\right)^2, \label{eq::Sigma}
\end{equation}
where $\omega_0$ is the characteristic frequency which is defined as
\begin{equation}
\omega_0 = \sqrt{\frac{8\gamma}{\rho  R_\mathrm{e}^3}}.
\end{equation}
Hereafter, we measure the length in the unit of the equatorial radius as $\bar{r} = r/R_\mathrm{e}$.
Equation~(\ref{eq::r_phi}) is then reduced to be
\begin{equation}
\frac{\phi}{\sqrt{1+\phi^2}} =-\bar{r} (1-\Sigma + \Sigma \bar{r}^2),
\end{equation}
or equivalently,
\begin{equation}
\phi = - \frac{g}{\sqrt{1 - g^2}} ,
\end{equation}
where
\begin{equation}
g(\bar{r}) \equiv \bar{r} (1 - \Sigma  + \Sigma \bar{r}^2).
\end{equation}
Finally, we obtain the form of the rotating droplet by the following integration, 
\begin{equation}
f(\bar{r}) = \int_0^{\bar{r}} \phi \diff \bar{r}. \label{eq::phi}
\end{equation}
While Eq.~(\ref{eq::phi}) can be expressed by the Jacobi elliptic functions,
the expression is inconvenient for comparison with numerical results.
We, therefore, derive the expression for the ratio of the equatorial radius
of the rotating droplet to the radius of the droplet at rest.
The volume of the droplet $V$ is expressed in terms of $f(\bar{r})$ by
\begin{equation}
V = 4 \pi R_\mathrm{e}^3 \int_0^1 \bar{r} f(\bar{r}) \diff \bar{r}.
\end{equation}
The integration by parts leads to
\begin{eqnarray}
\frac{V}{R_\mathrm{e}^3} &=&  4\pi \left[\frac{\bar{r}^2}{2} f \right]_0^1 - 2\pi \int_0^1 \bar{r}^2 \frac{\diff f}{\diff \bar{r}} \diff \bar{r}, \\
&=& 2\pi \int_0^1 \frac{\bar{r}^2g}{\sqrt{1-g^2}} \diff \bar{r},\\
&\equiv& 2\pi h,
\end{eqnarray}
where $h$ is the function of the reduced rotational frequency $\omega/\omega_0$.
Consider the sphere with the radius $R_0$ which volume is the same as the rotating droplet,
that is, $V \equiv 4 \pi R_0^3/ 3$.
Provided that the volume are not changed by rotation,
$R_0$ denotes the radius of the droplet at rest.
The radius ratio $R_\mathrm{e}/R_0$ is then expressed as,
\begin{equation}
R_\mathrm{e}/R_0 = \left( \frac{2}{3 h} \right)^{1/3}. \label{eq::scaling}
\end{equation}
Equation~(\ref{eq::scaling}) means that
deformation behaviors can be scaled for different rotation speed, volume of droplets, surface tension, and so forth.
Equation~(\ref{eq::scaling}) is shown as the solid line in Fig.~\ref{fig::diagram}.

A couple of things are worth to be noted.
First, the equilibrium form of the rotating droplet can be given by the
variational principle with appropriately chosen effective potential function, as described in Ref.~\cite{Brown}.
An Euler-Lagrange equation derived by the variational principle leads to
the Young-Laplace equation (\ref{eq::gamma}).
Therefore, the above arguments are completely equivalent to those from the
variational principle.
Second, one has to be careful with the definition of the characteristic
frequency defined in Eq.~(\ref{eq::Sigma}), since some researchers
take the radius of the droplet at rest $R_0$ as the scaling length,
while we use the equatorial radius of the rotating droplet $R_\mathrm{e}$.
Accordingly, the definition of $\Sigma$ and $\omega_0$ can be different for researchers.

\begin{figure}[tb]
\begin{center}
\includegraphics[width=0.6\linewidth]{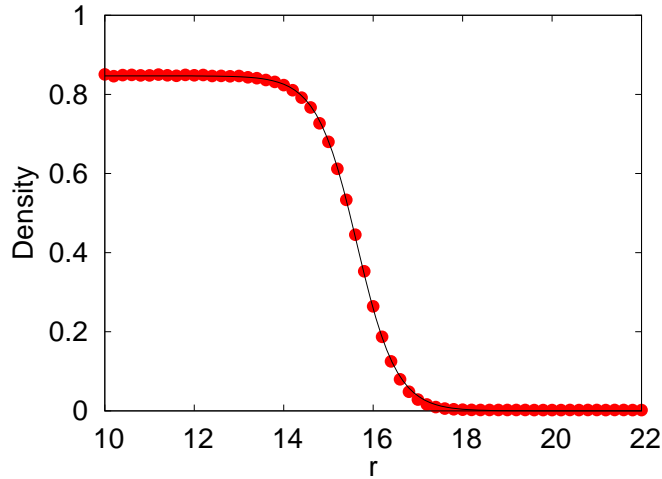}
\caption{The density of the droplet for $N=16000$.
The origin is set at the center of the droplet.
The solid line denotes the hyperbolic tangent function
of the form $\tanh((r-R_0)/\lambda)$ with the radius of the droplet $R_0 = 15.7(3)$
and the thickness of the surface $\lambda = 0.88(5)$.
The ratio of the thickness to the radius of the droplet is about $5.6\%$.
}
\label{fig::density}
\end{center}
\end{figure}

\section{Method} \label{sec::method}
In order to study the dynamics of rotating droplets, we perform MD simulations.
We use the truncated Lennard-Jones potential of the form
\begin{equation}  
V(r)=4\varepsilon\left[\left(\frac{\sigma}{r}\right)^{12}-\left(\frac{\sigma}{r}\right)^6+c_2\left(\frac{r}{\sigma}\right)^2+c_0\right],
\end{equation}
with the well depth $\varepsilon$, the atomic diameter $\sigma$, and the cut-off length $r_c$~\cite{cutoff_potential}.
The coefficients $c_0$ and $c_2$ are determined so that $V(r_c)=V'(r_c)=0$, i.e., the values of potential and the force become continuously zero at the truncation point.
We choose the cutoff-length as $r_c=3.0$.
In the following, we measure the physical quantities in the unit of 
the radius $\sigma$, the well depth $\varepsilon$, the Boltzmann constant $k_\mathrm{B}$, and the particle mass $m$.
Time step is chosen to be $0.005$.
The simulation box is the cube with the linear size $L$, and the periodic boundary condition is taken for all directions.
Total number of particles $N=4000, 8000$ and $16000$ are studied.
Temperature is controlled by Langevin thermostat~\cite{Langevin}.
Simulations are partially performed by MDACP which is freely available online~\cite{mdacp}

Particles are initially distributed spherically at the face-centered-cubic lattice, and time evolution is performed under the fixed temperature $T=0.6$.
The density of the total system is $\rho \equiv N/L^3 = 0.25$, which is the liquid-vapor coexistent phase at the temperature.
The system, therefore, reaches its equilibrium state after sufficiently long time, and the system contains a single droplet in the saturated vapor.
In the present study, $10^6$ steps are spent for the thermalization.
An typical snapshot of a droplet is shown in Fig.~\ref{fig::droplet}.
In order to identify the droplet, we define that two particles within the distance $1.3$ belong to the same cluster,
and the largest cluster is defined as the droplet which is shown in red particles in the following figures.
The density profile is shown in Fig.~\ref{fig::density}.
Since the interface between the liquid and the gas has finite thickness,
we have to define the equatorial radius to compare the results with the theory.
Here, we use the position of Gibbs surface to define the equatorial radius.
Let $r$ be distance from the center of a droplet.
The local density $\rho(r)$ is well approximated by the hyperbolic tangent function as,
\begin{equation}
\rho(r) = (\rho_\mathrm{L} - \rho_\mathrm{G}) \tanh \left( (r-R_\mathrm{G})/\lambda \right) + \rho_\mathrm{G}, \label{eq::tanh}
\end{equation}
with the liquid density $\rho_\mathrm{L}$, the gas density $\rho_\mathrm{G}$,
the position of Gibbs surface $R_\mathrm{G}$, and the thickness of the surface $\lambda$.
From the fitting using Eq.~(\ref{eq::tanh}), we determine the values of $\rho_\mathrm{L}$, $\rho_\mathrm{G}$, and $R_\mathrm{G}$.
Then we identify the position of Gibbs surface $R_\mathrm{G}$ as the equatorial radius $R_\mathrm{e}$.
We measure the surface tension of the droplets in this equilibrium state following the method
proposed by Ikeshoji {\it et al.}~\cite{Tamio}.
When the droplet is at rest, the equatorical radius $R_\mathrm{e}$ equals $R_0$.
The measured physical quantities of the droplet at rest are summarized in Table.~\ref{tbl::values}.

\begin{table}[tbp]
\begin{center}
\begin{tabular}{|c|c|c|c|c|}
\hline
$N$ & $R_0$ & $\rho$ & $\lambda$ & $\gamma$\\
\hline
4000 & 9.61(4) & 0.85(1) & 0.85(6) &0.62(1) \\
\hline
8000 & 12.4(3) & 0.85(1) & 0.88(5) &0.67(3) \\
\hline
16000 & 15.7(3) & 0.85(1) & 0.88(5) &0.72(1) \\
\hline
\end{tabular}
\end{center}
\caption{
Physical quantities of droplets at rest for $N=4000, 8000$, and $16000$.
$R_0$ is the position of Gibbs surface, $\rho$ is the density of the liquid,
$\lambda$ is the thickness of the surface, and $\gamma$ is the surface tension, respectively.
While the density and the thickness are almost independent of the size of the droplets,
the value of the surface tension increases as the size of the droplet increases.
}
\label{tbl::values}
\end{table}

\begin{figure*}[p]
\begin{center}
\includegraphics[width=0.7\linewidth]{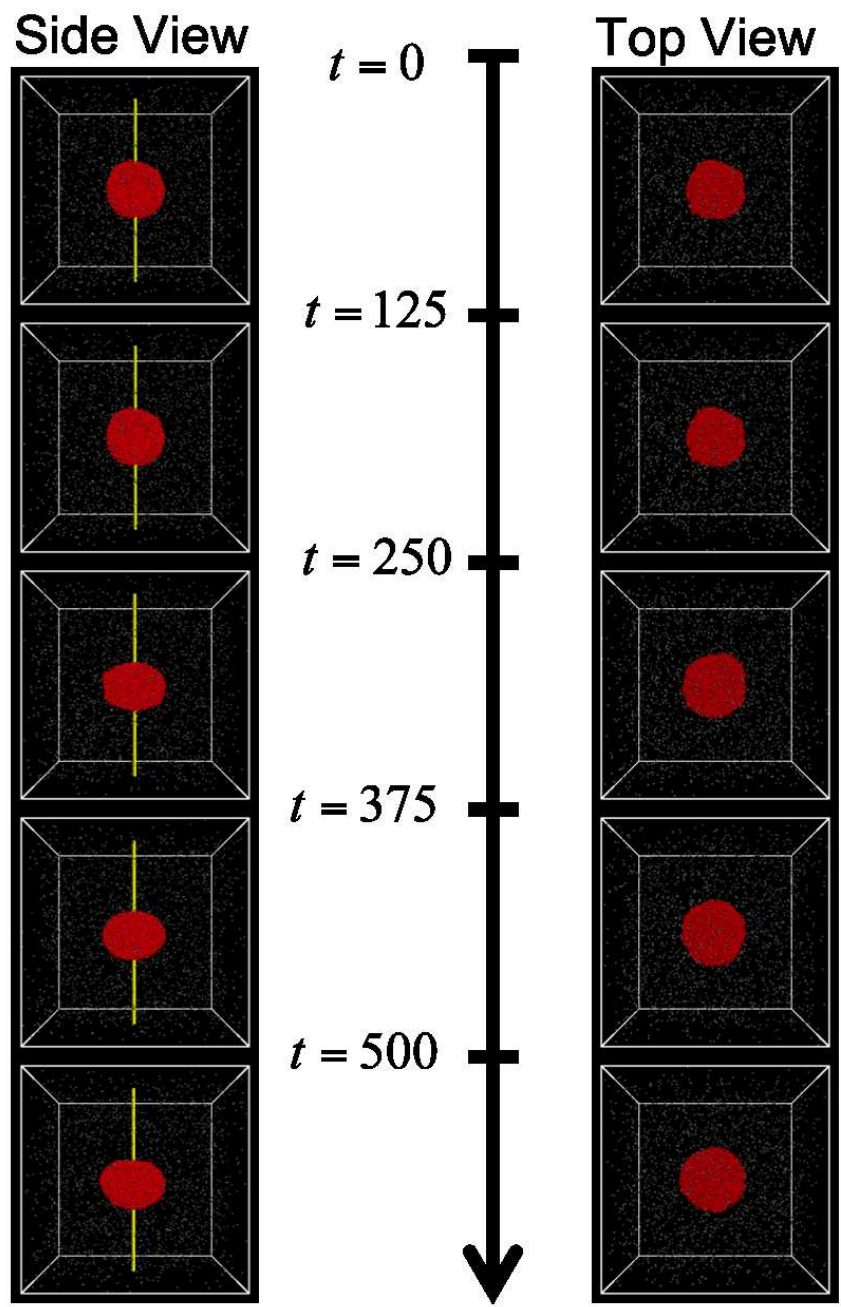}
\caption{
Time evolutions of a droplet of $N=16000$ for $\omega =0.025$.
The yellow line presents the axis of the rotation.
The views from the normal direction to the rotation axis are shown in the left,
and those from the parallel direction are shown in the right.
The form of the droplet changes from the sphere to the uniaxial.
}
\label{fig::om0025}
\end{center}
\end{figure*}

\begin{figure*}[p]
\begin{center}
\includegraphics[width=0.7\linewidth]{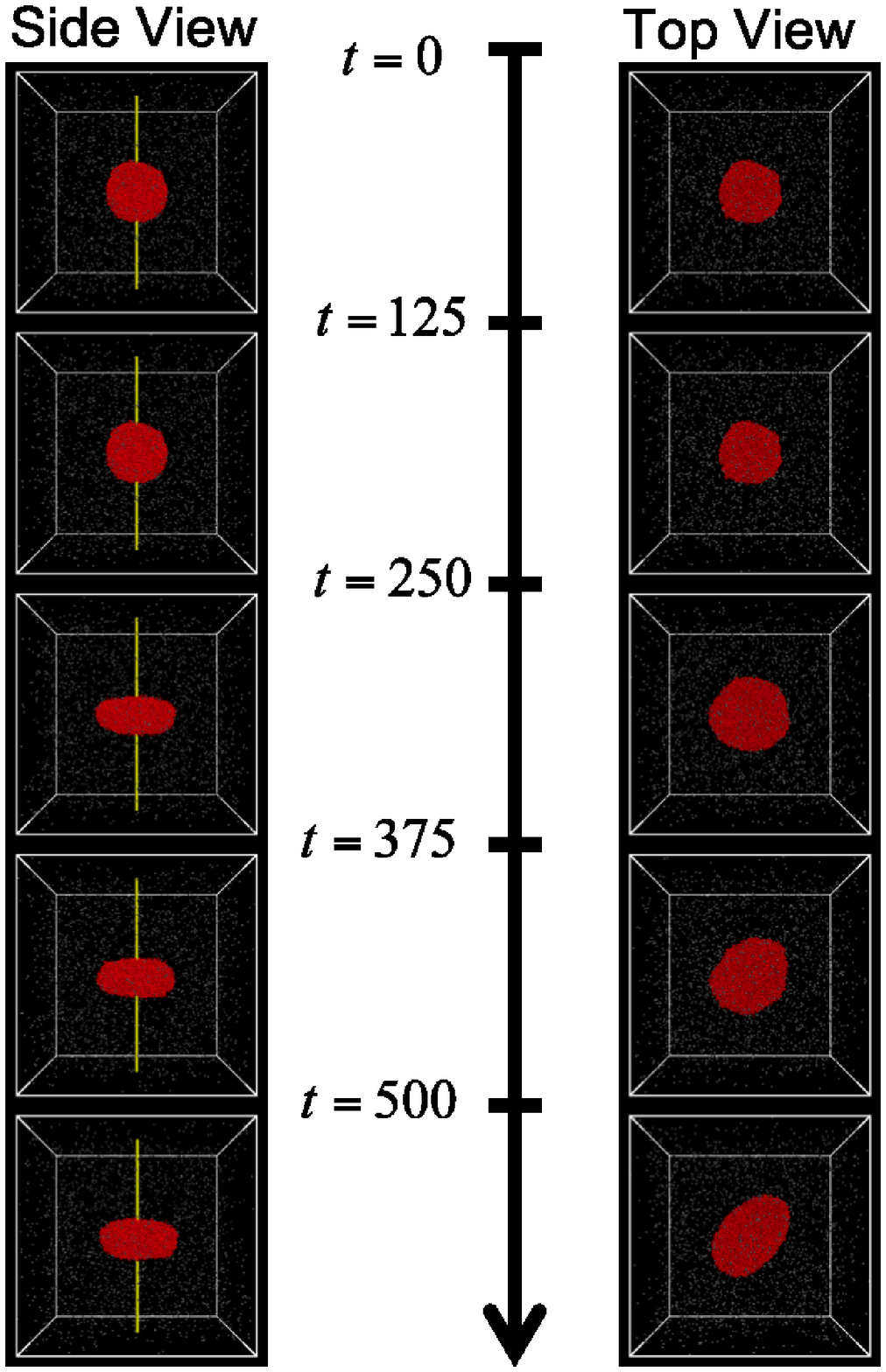}
\caption{
Same as Fig.~\ref{fig::om0025} for $\omega =0.05$.
The form of the droplet becomes biaxial.
}
\label{fig::om0050}
\end{center}
\end{figure*}

After the equilibrium droplet is obtained, the thermostat is turned off
and angular velocity $\omega$ is given to the droplet. 
As the droplet deforms, the moment of inertia also changes, and consequently, 
the angular velocity varies.
It makes difficult to investigate the relation between the angular velocity and the final form of the droplet precisely.
Therefore, we keep the angular velocity fixed throughout the time evolution
by the method similar to the velocity scaling scheme~\cite{Woodcock1971257}.
We observe the angular velocity of the droplet, and rescale the velocities
of the particles in the droplet so that the total angular velocity of the droplet
is kept to be the initial value.

\section{Results}

\subsection{Static Form of Rotating Droplets}

Two typical time evolutions of droplets are shown in Figs.~\ref{fig::om0025} and \ref{fig::om0050}.
The labels ``Side" and ``Top" in the figures denote that the view from the direction normal and parallel to the rotating axis, respectively.
Figure~\ref{fig::om0025} shows the results for the relatively slow rotation with $\omega = 0.025$.
The shape of the droplet viewed from the side becomes ellipse, while
the shape is kept circle viewed from the top. This implies that the droplet deforms to an oblate spheroid.
Figure~\ref{fig::om0050} shows the results for the larger angular velocity $\omega =0.05$.
The both shapes viewed from side and top become ellipses, \textit{i.e.},
the droplet becomes biaxial shape.
We find slight increase of temperature during rotation, but the increase is about $\sim 1 \%$.
Therefore, the influence of the heating by rotation can be negligible.

\begin{figure}[tb]
\begin{center}
\includegraphics[width=0.6\linewidth]{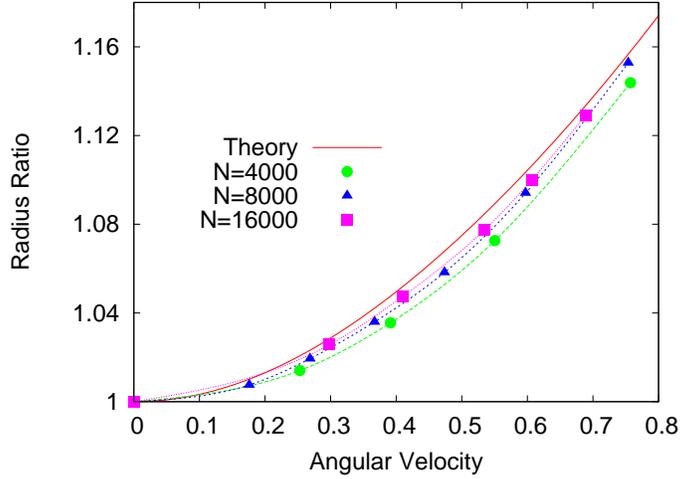}
\caption{The angular velocity dependence of the deformation of rotating droplets.
The radius ratio $R_\mathrm{e}/R_0$ is shown as a function of the reduced
angular velocity $\omega/\omega_0$.
The solid line denotes the theoretical prediction in Eq.~(\ref{eq::scaling}).
}
\label{fig::diagram}
\end{center}
\end{figure}

For the regime where the final shape of the droplet is uniaxial,
we compare the results from MD and the theoretical predictions.
We plot the ratio $R_\mathrm{e}/R_0$ as a function of the reduced frequency $\omega/\omega_0$ in Fig.~\ref{fig::diagram}.
While the deviations from the theoretical prediction are observed,
the difference between the numerical result and the theory becomes smaller
as the size of the droplet increases.
Note that, the experimental results such as in Ref.~\cite{PhysRevLett.56.452}
shows good agreements with the theretical prediction, which implies that
the size of the treated droplets is large enough so that the influence from the
thickness of the surface is negligibly-small.

\subsection{Deformation Dynamics}

\begin{figure}[htbp]
\begin{center}
\includegraphics[width=0.6\linewidth]{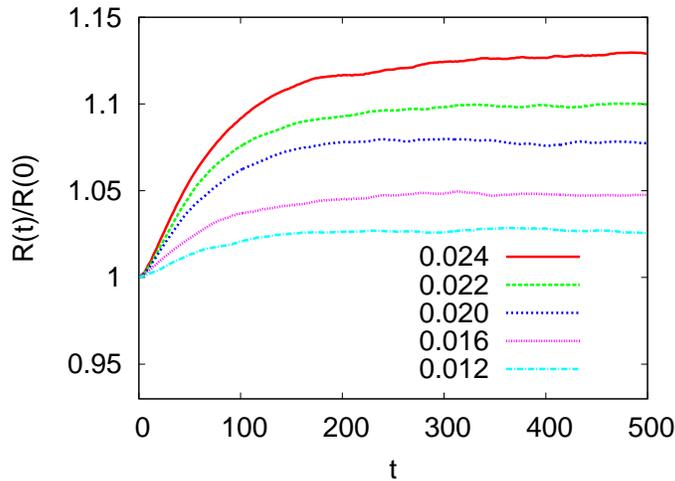}
\caption{Time evolutions of the deformation which is defined by the ratio of the current radius to the initial radius $R(t)/R(0)$
for different values of the angular velocities. Temperature is fixed to be $0.6$.
}
\label{fig::time_evolution_of_radius}
\end{center}
\end{figure}

\begin{figure}[htbp]
\begin{center}
\includegraphics[width=0.6\linewidth]{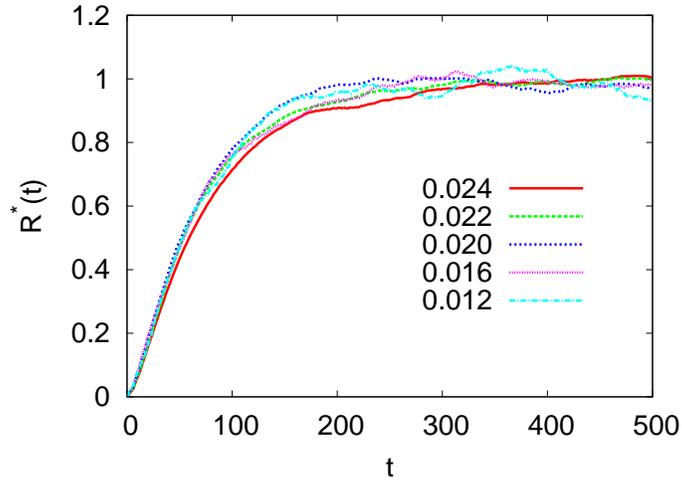}
\caption{
Time evolutions of reduced radius $R^*$ for different values of the angular velocity $\omega$.
They are well scaled. This implies that they share the identical timescale for their deformations,
while their amplitudes of the deformations are different.
}
\label{fig::scaled}
\end{center}
\end{figure}

Next, we observe the dynamics of the deformation.
While we have used the position of the Gibbs surface for the rotating droplet in steady state,
it is difficult to determine it when the droplet undergoes deformation, since 
the density profile of such cases cannot be determined accurately.
Therefore we measure the gyration radius instead of the position of the Gibbs surface in order to measure how a droplet deforms.
The gyration radius $R$ is defined as
\begin{equation}
R^2 \equiv \frac{1}{N} \sum r_i^2,
\end{equation}
where $r_i$ is the distance between particle $i$ to the rotating axis.
The time evolutions of the radius ratio $R(t)/R(0)$ for different values of $\omega$ are shown in Fig.~\ref{fig::time_evolution_of_radius}.
Deformation becomes larger for a larger value of $\omega$.
The time evolution of the radius is expected to be following exponential form as
\begin{equation}
R(t) =  R_\mathrm{eq} - \left[ R_\mathrm{eq} - R(0)\right] \exp(-t/\tau) , \label{eq::expform}
\end{equation}
with the final value of the radius $R_\mathrm{eq}$, i.e.,
$R_\mathrm{eq} \equiv \lim_{t\rightarrow \infty}  R(t)$.
We define the reduced radius $R^*$ as
\begin{equation}
R^*(t) = \frac{R(t)-R(0)}{R_\mathrm{eq} - R(0)},
\end{equation}
in order to investigate the characteristic timescale by removing the amplitude of the deformation.
The time evolutions of the reduced radii are shown in Fig.~\ref{fig::scaled}.
The data are collapsed into the single curve without scaling of the horizontal axis.
This implies that the deformations for different rotation velocity
have the same timescale. In other words, the timescale of the deformation
is almost independent of the angular frequency.

\begin{figure}[htbp]
\begin{center}
\includegraphics[width=0.6\linewidth]{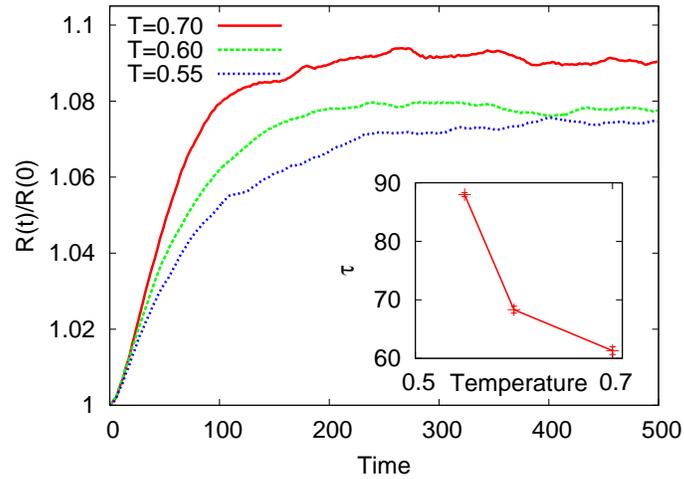}
\caption{Time evolutions of deformations for different temperatures.
The ratio of the radius $R(t)/R_0$ is shown, where $R(t)$ is the equatorial radius
and $R(0)$ is the initial radius, respectively.
(inset) The temperature dependence of the relaxation time $\tau$.
It shows that a droplet in higher temperature deforms quicker.
}
\label{fig::different_temperature}
\end{center}
\end{figure}

We also investigate influence of temperature.
Time evolutions of the radii of the droplets for different temperatures are shown in Fig.~\ref{fig::different_temperature}.
We estimate relaxation time $\tau$ defined in Eq.~(\ref{eq::expform}), and determine
the value of $\tau = 88.0(3)$, $68.3(6)$, and $61.3(7)$ for temperatures $T=0.55$, $0.60$, and $0.70$.
One can see that the timescale becomes shorter as temperature increases.
This means that the droplet deforms quickly for high temperature.

\section{Summary and Discussion}

To summarize, nonequilibrium deformation processes of rotating
droplets are investigated by molecular dynamics simulations.
First, we study the size-dependence of the deformations.
While the deformations of the rotating droplets in steady state are smaller 
than those of the theoretical predictions, the differences decrease as the size of the droplets increases.
The differences are the same order as the ratio of the thickness to the radius.
Therefore, the differences come from the fact that the surface of a droplet has finite thickness,
and the thickness cannot be ignored for small droplets.
This difference was not observed in the past studies such as in Ref.~\cite{PhysRevLett.56.452},
which implies that the treated droplet in the expriments are large enough.
Note that, the finite-size effect does not comes from the number of particles 
consisting the droplet, but the ratio of length of the interface to the radius of the droplet.
As the temperature increases, the length of the interface increases
and diverges at the critical point.
Therefore, the finite-size effect will be observed in the macroscopic experiments
in the region near the critical point.
For faster speed of rotation, the uniaxial form becomes unstable and 
biaxial form is observed as reported in the past study using the finite-element method~\cite{Brown}.

Nonequilibrium behaviors of the deformation are also observed.
We find that the rotation frequency only affects the form of the droplet
and does not change the timescale of deformations, when the rotation
is small enough so that the final shapes are uniaxial.
While the characteristic timescale is independent of the rotation speed,
it decreases as temperature increases.
In order to understand these nonequilibrium behaviors, we have constructed a following simple model.
Consider a spring-mass system in fluid.
A particle with mass $M$ is connected to the spring whose spring constant is $k$,
and the other end of the spring is fixed.
When the particle moves with the angular velocity $\omega$ around the fixed point,
then an overdumped equation of motion is written as,
\begin{equation}
\Gamma \frac{\diff x}{\diff t} = M(x_\mathrm{n} + x) \omega^2 - kx,
\end{equation}
where $x_\mathrm{n}$ is the natural length of the spring and $x$ is displacement from it.
$\Gamma$ denotes a phenomenological dissipation coefficient
which is proportional to viscosity of the fluid.
The solution of the equation is
\begin{equation}
x(t) =x_\mathrm{eq} - \left[x_\mathrm{eq} - x(0)\right] \exp(-t/\tau) \label{eq::dumpsolution}
\end{equation}
with the length at the equilibrium
\begin{equation}
x_\mathrm{eq} = \left[ \frac{M\omega^2}{k - M \omega^2} \right] x_\mathrm{n}
\end{equation}
and the relaxation time
\begin{equation}
\tau = \frac{\Gamma}{k - M\omega^2}. \label{eq_tau}
\end{equation}
The solution Eq.~(\ref{eq::dumpsolution}) corresponds to the behavior Eq.~(\ref{eq::expform}).
In this model, the spring constant $k$ corresponds to the surface tension $\gamma$,
which works as a restoring force (see Appendix).
When the rotational frequency is much smaller than the restoring force, i.e.,
$|\omega^2| \ll k/M$, then the characteristic timescale is reduced to be,
\begin{equation}
\tau = \frac{\Gamma}{k}. \label{eq_tau_approx}
\end{equation}
Then the timescale becomes independent of the rotational frequency, 
which is observed in the numerical simulations.
The temperature dependence of the timescale $\tau$ is not trivial,
since both restoring force $k$ and the dissipation coefficient $\Gamma$ depend on temperature.
Both surface tension and viscosity decreases as temperature increases, and
the observed deformation timescale becomes shorter as temperature increases.
This implies that the decrease of the viscosity is faster than that of the surface tension,
while situations may change for other types of liquid.
The dissipation coefficient $\Gamma$ depends not only on temperature,
but on size of a droplet. This size-dependence of dissipation is not trivial,
since the dissipation caused by deformation is difficult to be solved exactly.
These issues should be studied in the future.
With larger angular velocity, the deformation of the droplet becomes more complicated.
While rotation is almost rigid-body type for small angular velocity,
the internal flow of droplets can change the behavior as discussed in Ref.~\cite{Watanabe2009867}.
Non rigid-body rotation will involve dissipation, which causes change of temperature, etc.
Dynamics of such behaviors are also worth to be studied by molecular dynamics simulations.

\ack
The authors would like to thank Dr.~T. Shimada and Dr.~M. Suzuki for fruitful discussion.
This work is partially supported by Grants-in Aid for Scientific Research (Contracts No.~19740235), and KAUST GRB~(KUK-I1-005-04).

\appendix
\section{Restoring Force and Surface Tension}

In this appendix, we relate the phenomenological restoring force $k$ in Eq.~(\ref{eq::dumpsolution})
to the surface tension $\gamma$ and the rotating frequency $\omega$.
Consider a rotating droplet consisting $N$ particles.
When the droplet rotate slow enough, the form of the droplet is uniaxial and
well approximated by an oblate spheroid.
The polar radius $R_\mathrm{p}$ and the equatorial radius $R$
satisfy the following equation,
\begin{equation}
N = \frac{4\pi}{3}  \rho R^2 R_\mathrm{p}.
\end{equation}
The surface area $S$ of this droplet is
\begin{equation}
S = 2 \pi \left( R^2 +\displaystyle \frac{R_\mathrm{p}^2 \tanh^{-1} e }{e} \right),
\end{equation}
with the eccentricity $e \equiv \sqrt{1 - R_\mathrm{p}^2/R^2} $.
When the rotation velocity is small enough, the eccentricity is very close to unity.
Then the approximation $\tanh^{-1} e \sim e$ for $|e| \ll 1$ leads to 
\begin{equation}
S = 2 \pi ( R^2 + R_\mathrm{p}^2).
\end{equation}
The moment of inertia $I$ of the droplet is 
\begin{equation}
I = \frac{2}{5}mNR^2,
\end{equation}
where $m$ is the mass of the particles.
Consequently, the effective potential of the droplet can be written as
\begin{eqnarray}
U &=& S \gamma - \frac{1}{2}I \omega^2 \\
&=& \left( 2\pi \gamma - \frac{mN \omega^2}{5} \right) R^2
+ \frac{9N^2\gamma}{8\pi \rho^2 R^4},
\end{eqnarray}
with the rotating frequency $\omega$ and the surface tension $\gamma$.
The contribution from the inner pressure is ignored here,
since the observed pressure is almost constant during its rotation in the numerical results.
The restoring force for the form of droplets $F(r)$ is written as
\begin{eqnarray*}
F(r) &=& - \displaystyle \frac{\diff U}{\diff R}\\
&=&  -\left( 4\pi \gamma - \frac{2mN\omega^2}{5} \right) R
+ \frac{9N^2\gamma}{2\pi \rho^2 R^5}.
\end{eqnarray*}
The mechanical equilibrium radius is given by $F(R_\mathrm{eq}) = 0$, which is
\begin{equation}
R_\mathrm{eq} = \left[  \displaystyle\frac{45N^2\gamma}{2 \pi \rho^2 (20 \pi \gamma - 2 mN^2\omega^2)}  \right]^{1/6}.
\end{equation}
The linear approximation around the mechanical equilibrium point gives the 
effective spring constant of the droplet $K$ as
\begin{equation}
K \equiv - \left. \frac{\diff F}{\diff R}\right|_{R = R_\mathrm{eq}}  = \left( 24\pi \gamma - \frac{12mN \omega^2}{5} \right). \label{eq_k}
\end{equation}
This $K$ corresponds to the restoring force $k-M\omega^2$ in Eq.~(\ref{eq_tau}).
Then identifying $mN$ with $M$, we have the characteristic timescale of this system as
\begin{equation}
\tau = \displaystyle \frac{\Gamma}{24\pi \gamma - 12 M \omega^2/5}.
\end{equation}
When the rotation of the droplet is slow enough, the relaxation time becomes
\begin{equation}
\tau = \displaystyle \frac{\Gamma}{24\pi \gamma},
\end{equation}
which is essentially equivalent to Eq.~(\ref{eq_tau_approx}).


\section*{References}


\begin{thebibliography}{10}
\bibitem{spray_combustion}
Sirignano W A 1983
{\it Progress in Energy and Combustion Science} \textbf{9} 291

\bibitem{inkjet}
Calvert P 2001
\newblock {\it Chemistry of Materials} \textbf{13} 3299

\bibitem{sputtering}
Bringa E M,  Johnson R E, and Jakas M 1999
\newblock {\it Phys. Rev. B} \textbf{60} 15107

\bibitem{Plateau}
Plateau J 1863
\textit{Annual Report of the Board of Regents of the Smithsonian Institution, Washington DC, 1863} (Washington DC: The Smithsonian Institution) 207--285

\bibitem{PhysRevLett.101.234501}
Hill R J A and Eaves L 2008
\textit{Phys. Rev. Lett.} \textbf{101} 234501

\bibitem{Brown}
Brown R A and Scriven L E1980
\textit{Proc. R. Soc. Lond. A} \textbf{371} 331

\bibitem{Watanabe2009867}
Watanabe T 2009
\textit{Physics Letters A} \textbf{373} 867

\bibitem{PhysRevLett.56.452}
Wang T G, Trinh E H, Croonquist A P, and Elleman D D 1986
{\it Phys. Rev. Lett.} \textbf{56} 452

\bibitem{Chandrasekhar}
Chandrasekhar S 1965
{\it Proc. R. Soc. Lond. A.} \textbf{286} 1

\bibitem{cutoff_potential}
Stoddard S D and Ford J 1973
{\it Phys. Rev. A} \textbf{8} 1504

\bibitem{Langevin}
Adelman S A and Doll J D 1976
{\it J. Chem. Phys.} \textbf{64} 2375

\bibitem{mdacp}
\verb|http://mdacp.sourceforge.net/|

\bibitem{Tamio}
Ikeshiji T, Hafskjold B, and Furholt H 2003
{\it Molecular Simulation} \textbf{29} 101

\bibitem{Woodcock1971257}
Woodcock L V 1971
{\it Chem. Phys. Lett.} \textbf{10} 257
\end{thebibliography}
\end{document}